\documentclass[twocolumn,showpacs,prc]{revtex4}
\usepackage{graphicx}
\usepackage{dcolumn}
\usepackage{bm}
\begin{document}

\title{Photo-Disintegration of the Iron Nucleus\\ 
in Fractured Magnetite Rocks with  Magnetostriction}
\author{A. Widom and J. Swain}
\affiliation{Physics Department, Northeastern University, Boston MA USA}
\author{Y.N. Srivastava}
\affiliation{Physics Department \& INFN, University of Perugia, Perugia IT}

\begin{abstract}
There has been considerable interest in recent experiments  
on iron nuclear disintegrations observed when rocks containing such 
nuclei are crushed and fractured. The resulting nuclear 
transmutations are particularly strong for the case of magnetite 
rocks, i.e. loadstones. We argue that the fission of the iron 
nucleus is a consequence  of photo-disintegration. The electro-strong 
coupling between electromagnetic fields and nuclear giant dipole 
resonances are central for producing observed nuclear reactions. 
The large electron energies produced during the fracture of 
piezomagnetic rocks are closely analogous to the previously 
discussed case of the fracture of piezoelectric rocks. In both cases 
electro-weak interactions can produce neutrons and neutrinos from 
energetic protons and electrons thus inducing nuclear transmutations. 
The electro-strong condensed matter coupling discussed herein represents 
new many body collective nuclear photo-disintegration effects.
\end{abstract}


\maketitle

\section{Introduction \label{intro}}

Recent measurements of nuclear 
reactions\cite{Carpinteri:2010,Carpinteri:2011,Manuello:2011} 
that accompany the 
fracturing\cite{Chiodoni:2011,Carpinteri:2012,Lacidogna:2012,Borla:2012} 
of piezoelectric and piezomagnetic rocks have inspired a great deal 
of interest. The contribution of electro-weak processes to the production 
of neutrons, 
\begin{equation}
e^- + p^+ \to n + \nu_e ,
\label{intro1}
\end{equation}
during the fracture of piezoelectric rocks has been discussed in previous 
work\cite{Widom:2013}. One of the purposes of this work is to expand the 
theory to include the fracture of rocks\cite{Koshevaya:2008} with  
magnetostriction, again employing the conversion of mechanical energy 
(phonons) to electromagnetic energy (photons) and vice-versa.

\begin{figure}
\scalebox {0.6}{\includegraphics{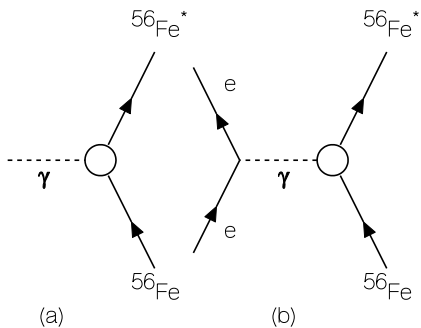}}
\caption{Shown in (a) is the Feynman diagram for producing an excited 
compound nucleus from a real Bremsstrahlung photon. Shown in (b) 
is a similar process due to an incident electron exchanging a virtual 
photon with the nucleus.} 
\label{fig1}
\end{figure}

Our purpose is also to investigate other sources of nuclear radiation, in 
particular those radiations that are a consequence of 
electro-strong\cite{Swain:2013} 
nuclear fission reactions. To see what is involved, consider the fracture 
of \begin{math} Fe_3O_4 \end{math} magnetite rocks, i.e. magnetic 
semiconductor loadstones. The fracture process accelerates electrons which
in turn produce electromagnetic radiation. Such radiation can induce 
the photo-disintegration of the iron nucleus. The absorption of a photon 
can cause a transition to an excited compound nuclear state, e.g.  
\begin{equation}
\gamma +\ ^{56}Fe \to \ ^{56}Fe^* . 
\label{intro2}
\end{equation}
The quantum electrodynamic excitation to a compound nuclear state is 
exhibited in the Feynman diagrams of FIG. \ref{fig1}. The photon nuclear 
vertex coupling is via the iron nucleus giant dipole resonance\cite{Snover:1986}. 
The strong interactions control the decay of the compound nucleus 
\begin{equation}
 ^{56}Fe^*\ \to ({\rm fission\ products}). 
\label{intro3}
\end{equation}
In general terms, the photo-disintegration of nuclei is presently 
well understood. 

In Sec.\ref{gdr} the giant dipole resonant coupling to the electromagnetic 
field is reviewed. The total cross section for absorbing the photon in 
FIG. \ref{fig1} (a) has a peak value of 
\begin{equation}
\sigma_0=4\pi Z \alpha \left(\frac{\hbar }{M\Gamma }\right),
\label{intro4}
\end{equation}
wherein \begin{math} Z=26  \end{math} is the number of protons in the 
iron nucleus, the quantum electrodynamic coupling strength is 
\begin{math} \alpha = (e^2/\hbar c)\approx 1/137.036 \end{math}, 
the proton mass is \begin{math} M \end{math} and the fission decay rate 
of the compound nucleus in Eq.(\ref{intro3}) is \begin{math} \Gamma \end{math}.
The silicon and aluminum photo-disintegration fission channels are briefly
discussed in Sec.\ref{ifc}.

In Sec.\ref{rf} we consider the retarding force 
\begin{equation}
\bar{F}=-\frac{dE}{dx}\ ,
\label{intro5}
\end{equation}
or energy loss per unit length for an energetic electron passing through 
magnetite. The ratio of how much energy \begin{math} dE \end{math} is 
lost to other electrons in atomic transitions and how much energy 
\begin{math} d\tilde{E} \end{math} is lost to giant dipole excitations 
of the compound nucleus is computed in detail. Formally, the energy 
transfer efficiency \begin{math} \eta =(d\tilde{E}/dE)\sim 1\% \end{math}.

In Sec.\ref{pme} a theoretical explanation is provided for the experimental 
evidence that fracturing loadstones produces photo-disintegration 
fission products of \begin{math} ^{56}Fe \end{math} nuclei. The elastic 
energy of mesoscopic microcrack production during a fracture ultimately yields 
a major macroscopic fracture separation. The mechanical energy is converted by  
magnetostriction into electromagnetic field energy. The electromagnetic 
field energy decays via radio frequency (microwave) oscillations. The 
radio frequency  fields accelerate the condensed matter electrons which then 
collide with nuclei producing fission products on the surfaces of these 
microcracks. In the concluding Sec.\ref{conc}, the nature of fission microcrack 
wall remnants are discussed.

\section{Giant Dipole Resonance\label{gdr}}

Let \begin{math} \left|0 \right> = \left|^{56} Fe\right> \end{math} represent the 
ground state internal wave function of the iron nucleus. Representing the dipole 
approximation for the interaction in FIG. \ref{fig1}, 
\begin{equation}
H_{\rm int}=-{\bf E}\cdot {\bf d},
\label{gdr1}
\end{equation}
the nuclear polarizability 
\begin{equation}
\beta(\zeta )=\frac{i}{3\hbar} \int_0^\infty e^{i\zeta t} 
\left<0\right|{\bf d}(t)\cdot {\bf d}(0)-
{\bf d}(0)\cdot {\bf d}(t) \left|0\right>dt. 
\label{gdr2}
\end{equation}
The ground state of the \begin{math} ^{56} Fe \end{math} nucleus has zero 
spin. The polarizability is thereby an isotropic tensor.

\subsection{Photon Cross Sections \label{pcs}}

The elastic photon scattering amplitude for 
\begin{equation}
\gamma + \ ^{56} Fe \to \gamma + \ ^{56} Fe
\label{gdr3}
\end{equation}
is given by 
\begin{equation}
{\cal F}_{fi}(\omega )=\left(\frac{\omega }{c}\right)^2 
\beta (\omega +i0^+){\bf e}_f^* \cdot {\bf e}_i
\label{gdr4}
\end{equation}
wherein \begin{math} {\bf e}_{i,f}  \end{math} are, respectively, 
the initial and final photon polarization vectors and 
\begin{math} \omega \end{math} is the photon frequency. The elastic 
photon cross section is thereby  
\begin{eqnarray}
\sigma_{\rm el}(\omega )=\frac{1}{2}\sum_f \sum_i 
\int \left| {\cal F}_{fi}(\omega ) \right|^2 d\Omega_f\ ,
\nonumber \\ 
\sigma_{\rm el}(\omega )=
\frac{8\pi }{3} \left(\frac{\omega }{c} \right)^4 
\left|\beta (\omega +i0^+)\right|^2 .
\label{gdr5}
\end{eqnarray}
wherein we have averaged over initial polarization states and summed over final 
polarization states. The total cross section follows from the optical theorem 
\begin{eqnarray}
\sigma_{\rm tot}(\omega )=\left(\frac{4\pi c}{\omega }\right)
{\Im m}{\cal F}_{ii}(\omega ),
\nonumber \\ 
\sigma_{\rm tot}(\omega )=
\left(\frac{4\pi \omega}{c}\right){\Im m}\beta (\omega +i0^+).
\label{gdr6}
\end{eqnarray}
The inelastic cross section \begin{math} \sigma_{\rm in}(\omega ) \end{math} 
for the central reaction 
\begin{equation}
\gamma +\ ^{56}Fe \to \ ^{56}Fe^* \to ({\rm fission\ products}) 
\label{gdr7}
\end{equation}
follows from Eqs.(\ref{gdr5}) and (\ref{gdr6}) via the sum rule  
\begin{equation}
\sigma_{\rm tot}(\omega ) = 
\sigma_{\rm el}(\omega )+\sigma_{\rm in}(\omega ). 
\label{gdr8}
\end{equation}

\subsection{Dispersion Relations \label{dr}}

From Eq.(\ref{gdr2}), it is expected that the nuclear polarizability 
obey a dispersion relation for \begin{math} {\Im m}\ \zeta > 0 \end{math} 
of the form 
\begin{equation}
\beta(\zeta ) = 
\frac{2}{\pi } \int_0^\infty 
\left[\frac{\omega {\Im m}\beta (\omega +i0^+)d\omega }{\omega ^2 - \zeta ^2}\right].   
\label{gdr9}
\end{equation}
With 
\begin{equation}
\hbar \omega_{n0}=E_n-E_0\ \ \ {\rm and}
\ \ \ {\bf d}_{n0}=\left<n\right|{\bf d}\left|0\right>,
\label{gdr9me}
\end{equation}
we have 
\begin{equation}
{\Im m}\beta (\omega +i0^+)=\frac{\pi }{3\hbar }
\sum_n |{\bf d}_{n0}|^2 \delta (\omega -\omega_{n0})
\label{gdr9im}
\end{equation}
For example, the static polarizability of the nucleus is given by 
\begin{equation}
\beta =\lim_{\zeta \to i0^+} \beta(\zeta ) = 
\frac{2}{\pi } \int_0^\infty {\Im m}\beta (\omega +i0^+)
\frac{d\omega }{\omega }\ .  
\label{gdr9s}
\end{equation} 
Employing Eq.(\ref{gdr1}) in second order perturbation theory 
\begin{eqnarray}
\frac{1}{2}\beta |{\bf E}|^2=
\sum_n \frac{\left|\left<n\right|H_{\rm int} \left|0\right>\right|^2}{E_n-E_0}\ ,
\nonumber \\ 
\beta =\frac{2}{3\hbar }
\sum_n \frac{\left| {\bf d}_{n0} \right|^2}{\omega_{n0}}\ .
\label{gdr9pt}
\end{eqnarray} 
For a nucleus of charge \begin{math} Z \end{math}, the equal time commutation 
relation 
\begin{equation}
\frac{i}{3\hbar }
\left(\dot{\bf d}\cdot {\bf d}-{\bf d}\cdot \dot{\bf d} \right) 
=\frac{Ze^2}{M},
\label{gdr10}
\end{equation}
wherein \begin{math} M \end{math} is the proton mass, yields in virtue of 
Eqs.(\ref{gdr2}) and (\ref{gdr10}) the large frequency limit 
\begin{equation}
\lim_{|\zeta|\to \infty} \zeta^2 \beta (\zeta )
=-\left(\frac{Ze^2}{M}\right). 
\label{gdr11}
\end{equation}
Eqs.(\ref{gdr9}) and (\ref{gdr11}) imply the sum 
rule\cite{Thomas:1925,Kuhn:1925,Reiche:1925}
\begin{equation}
\frac{2}{\pi }\int_0^\infty \omega {\Im m}\beta (\omega +i0^+)d\omega  
=\frac{Ze^2}{M}\ , 
\label{gdr12}
\end{equation}
or equivalently 
\begin{equation}
\int_0^\infty \sigma_{\rm tot}(\omega )d\omega = 
2\pi^2\left(\frac{Ze^2}{Mc}\right)
=2\pi^2 Z\alpha \left(\frac{\hbar}{M}\right). 
\label{gdr13}
\end{equation}
The above dispersion relation and sum rule are implemented in simple 
nuclear giant dipole resonant models. 

\subsection{Single Giant Resonance Model \label{srm}}

\begin{figure}
\scalebox {0.6}{\includegraphics{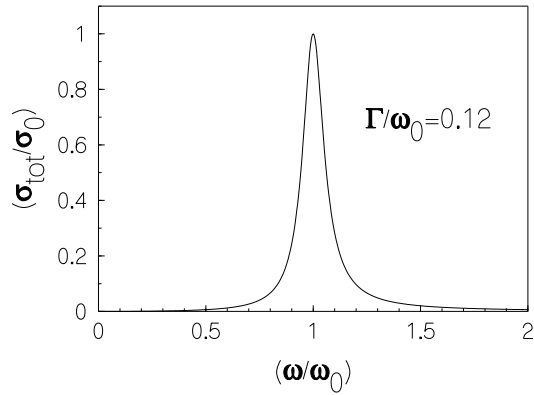}}
\caption{Shown is the total cross section employing a single 
nuclear dipole giant resonance model according to 
Eq.(\ref{gdr15}). The resonant energy for creating a compound 
nucleus is $\hbar \omega_0$ as in Eq.(\ref{intro2}). The transition 
rate is $\Gamma $ for the compound nucleus to decay into fission 
products as in Eq.(\ref{intro3}). The peak cross section is 
$\sigma_0=4\pi Z\alpha \hbar /M\Gamma$. The resonant width 
$\Gamma \approx 0.12\ \omega_0$ for $^{56} Fe$ is only approximate.} 
\label{fig2}
\end{figure}

The most simple model for discussing the nuclear giant dipole 
resonance involves a single damped harmonic oscillator with 
resonant frequency \begin{math} \omega_0 \end{math} and damping 
coefficient \begin{math} \Gamma \end{math}.
\begin{equation}
\beta (\zeta )=
\frac{Ze^2}{M\big(\omega_0^2-\zeta^2-i\zeta \Gamma \big)}\ .
\label{gdr14}
\end{equation}
Eqs.(\ref{gdr6}) and (\ref{gdr14}) then imply a total cross section 
\begin{eqnarray}
\sigma_{\rm tot}(\omega )=\left(\frac{4\pi Ze^2}{Mc}\right)
\left[\frac{\omega ^2 \Gamma }{(\omega_0^2-\omega^2)^2+\omega^2 \Gamma^2 } \right],
\nonumber \\ 
\sigma_{\rm tot}(\omega )=\sigma_0
\left[\frac{\omega ^2 \Gamma ^2}{(\omega_0^2-\omega^2)^2+\omega^2 \Gamma^2 } \right],
\nonumber \\ 
\sigma_0=\left(\frac{4\pi Ze^2}{Mc\Gamma }\right)=
4\pi Z\alpha \left(\frac{\hbar }{M\Gamma }\right).
\label{gdr15}
\end{eqnarray}
The single resonance Eq.(\ref{gdr15}) is plotted in FIG. \ref{fig2}. 
 
\subsection{Model Parameters \label{mp}}

In the Migdal theory\cite{Migdal:1967,Migdal:1990} of the 
nuclear Landau-Fermi liquid, an isotopic spin zero sound mode of 
velocity \begin{math} c_0 \end{math} confined to a spherical 
cavity has a frequency 
\begin{equation}
\omega_0=\frac{zc_0}{R}\ \ 
({\rm with\ the\ lowest\ root\ of}\ \ \ z=\tan z>0).
\label{gdr16}
\end{equation}
The radii of nuclei obey 
\begin{equation}
R\approx r_0 A^{1/3}\ \ {\rm wherein}
\ \ r_0\approx 1.2\ {\rm fm}.
\label{gdr17}
\end{equation}
With the ratio of zero sound speed to light speed given by 
\begin{math} (c_0/c)\approx 0.1  \end{math} we then have the 
prediction
\begin{equation}
\hbar \omega_0\approx \frac{70\ {\rm MeV}}{A^{1/3}}
\ \ \Rightarrow \ \ \hbar \omega_0(^{56}Fe)\approx 20\ {\rm MeV}.
\label{gdr18}
\end{equation}
Since the Landau-Fermi liquid theory gives rise to the complex zero sound 
mode frequency \begin{math} \varpi =c_0 k-(i/2)Dk^2 \end{math} as 
\begin{math} k\to 0 \end{math}, the width of the zero sound resonance 
must have the form 
\begin{equation}
\Gamma =D\left(\frac{\omega_0}{c_0}\right)^2=\frac{z^2D}{R^2}=\frac{\Gamma_0}{A^{2/3}}
\label{gdr19}
\end{equation}
wherein 
\begin{equation}
\Gamma_0=\left(\frac{z^2D}{r_0^2}\right). 
\label{gdr20}
\end{equation}
Thus, 
\begin{equation}
\frac{\Gamma}{\omega_0}=\left(\frac{zD}{c_0r_0}\right)\frac{1}{A^{1/3}}\approx 
\frac{0.5}{A^{1/3}}\ .
\label{gdr21}
\end{equation}
The above estimates are semi-quantitative but to the best of our knowledge are 
new.

\section{Disintegration Fission Channels \label{ifc}}

In the photo-disintegration of iron 
\begin{eqnarray}
\gamma +\ ^{56}Fe \to \ ^{56}Fe^* 
\to ({\rm fission\ products}). 
\label{ifc1}
\end{eqnarray}
one should consider the silicon and aluminum fission products 
within the prominent decay channels.
If in accordance with the liquid drop model, the decay of 
of the compound excited nucleus \begin{math} ^{56}Fe^*  \end{math}
is preferentially into two equal excited aluminum droplets, the decay reads 
\begin{equation}
^{56}Fe^* \to 2\ \ ^{28}Al^* .  
\label{ifc2}
\end{equation}
Each of the resulting aluminum nuclei, then undergoes the weak decay  
\begin{equation}
 ^{28}Al^* \to \  ^{28}Si +e^- +\bar{\nu}_e.  
\label{ifc3}
\end{equation}
Altogether, 
\begin{equation}
\gamma +\ ^{56}Fe \to 2 \ \ ^{28}Si +2e^- +2\bar{\nu}_e 
\ \ ({\rm two\ Si\ channel}). 
\label{ifc4}
\end{equation}
If two neutrons evaporate from each of the two resulting aluminum nuclei 
\begin{equation}
 ^{28}Al^* \to \  ^{27}Al + n,  
\label{ifc5}
\end{equation}
then a possible channel is 
\begin{equation}
\gamma +\ ^{56}Fe \to 2 \ \ ^{27}Al +2n  
\ \ ({\rm two\ Al\ channel}). 
\label{ifc6}
\end{equation}
A third channel is evidently 
\begin{eqnarray}
\gamma +\ ^{56}Fe \to  \ ^{27}Al +\ ^{28}Si 
+ n +e^- +\bar{\nu}_e\ ,
\nonumber \\   
\ \ ({\rm Al+Si\ channel}). 
\label{ifc7}
\end{eqnarray}
In terms of the branching ratio,
\begin{equation}
b =\frac{\Gamma \Big(\  ^{28}Al^* \to \  ^{28}Si +e^- +\bar{\nu}_e  \Big)}
{\Gamma \Big( \ ^{28}Al^* \to \  ^{27}Al + n \Big)}\ ,
\label{ifc8}
\end{equation}
one should be able to compute the branching ratio of the above three fission 
products. 

In magnetite \begin{math} Fe_3 O_4 \end{math}, there will also be the 
photo-disintegration of the \begin{math} ^{16} O  \end{math} nucleus,  
\begin{equation}
\gamma +\ ^{16}O\to \  ^{16}O^* \to ({\rm fission\ products}).
\label{ifc9}
\end{equation}
Possible fission channels are thereby 
\begin{eqnarray}
^{16}O^* \to \ ^{15}O+n , 
\nonumber \\ 
^{16}O^* \to \ ^{15}N+ p^+ , 
\nonumber \\ 
^{16}O^* \to \ ^{14}N+ d^+ , 
\nonumber \\ 
^{16}O^* \to \ ^{13}N+ t^+ , 
\nonumber \\ 
^{16}O^* \to \ ^{13}C +\ ^{3}He ,
\nonumber \\ 
^{16}O^* \to \ ^{12}C +\alpha ,
\nonumber \\ 
^{16}O^* \to \ ^{10}B + \ ^6 Li . 
\label{ifc10}
\end{eqnarray}
The photo-disintegration of oxygen gives rise to diverse types 
of nuclear radiation.

\section{Retardation Forces \label{rf}}

Consider a beam of electrons each having an energy 
\begin{equation}
E=mc^2\gamma =\frac{mc^2}{\sqrt{1-(v/c)^2}} \ .
\label{rf1}
\end{equation}
passing through magnetite. The energy loss per unit length for 
the electron is the retarding force 
\begin{equation}
F=-\frac{dE}{dx}\ .
\label{rf2}
\end{equation}
If the fast electron causes other electrons to undergo atomic transitions, 
then the retarding force takes on the well known form\cite{Berestetskii:1990}
\begin{equation}
F=\left(\frac{4\pi ne^4}{mv^2}\right)
\left[\ln\left(\frac{mv^2\gamma^2}{\hbar \omega_i}\right)-
\left(\frac{v}{c}\right)^2 \right],
\label{rf3}
\end{equation}
wherein \begin{math} \hbar \omega_i  \end{math} is the log mean ionization 
energy of an atom and \begin{math} n \end{math} is the density of 
electrons per unit volume. 

For \begin{math} \tilde{n}  \end{math} iron \begin{math}\ ^{56}Fe \end{math} 
nuclei per unit volume, the retardation force due to the giant dipole 
resonance at frequency \begin{math} \omega_0  \end{math} in iron is given by  
\begin{equation}
\tilde{F}=\left(\frac{4\pi \tilde{n} Z^2e^4}{Mv^2}\right)
\left[\ln\left(\frac{Mv^2\gamma^2}{\hbar \omega_0}\right)-
\left(\frac{v}{c}\right)^2 \right],
\label{rf4}
\end{equation}
wherein \begin{math} M \end{math} is the proton mass.

The photo-disintegration efficiency may be defined as 
\begin{equation}
\eta = \left(\frac{\tilde{F}}{F}\right)=\left(\frac{d\tilde{E}}{dE}\right) ,  
\label{rf5}
\end{equation}
wherein \begin{math} d\tilde{E} \end{math} is the energy transferred from 
a fast electron to fission products and \begin{math} dE \end{math} is the 
energy transferred from a fast electron to atomic electronic transitions. 
The ratio of the energy excitations is small 
\begin{math} (\omega_i/\omega_0)\sim 10^{-7}  \end{math}.
Eqs.(\ref{rf3}), (\ref{rf4}) and (\ref{rf5}) imply for 
\begin{math} \gamma\gg 1 \end{math} 
\begin{equation}
\eta \approx \left(\frac{Z^2\tilde{n}m}{nM}\right)
\left\{\frac{\ln [({Mv^2\gamma^2}/{\hbar \omega_0})]-
(v/c)^2}{\ln [({mv^2\gamma^2}/{\hbar \omega_i})]-
(v/c)^2}\right\}.   
\label{rf6}
\end{equation}
Typically, the nuclear yield is as expected of the order of 
\begin{math} \eta \sim 1\%  \end{math}. 

\section{Fractured Magnetic Rocks \label{pme}}

The thermodynamic properties of magnetite \begin{math} Fe_3O_4 \end{math} 
are determined by the energy per unit volume 
\begin{math} u=u(s,{\bf w},{\bf H})  \end{math} wherein 
\begin{math} s \end{math} is the entropy per unit volume, 
\begin{math} {\bf w} \end{math} is the strain and 
\begin{math} {\bf H} \end{math} is the magnetic intensity  
\begin{equation}
du=Tds+{\bf \sigma}:d{\bf w}-{\bf M}\cdot d{\bf H}.
\label{pme1}
\end{equation} 
Magnetite at room temperature is a ferromagnet. Within a magnetic domain, 
one may formally define a piezomagnetic coefficient by the thermodynamic 
derivatives
\begin{equation}
\beta_{i,jk}=\left(\frac{\partial M_i}{\partial w_{jk}}\right)_{s,{\bf H}} 
=-\left(\frac{\partial \sigma_{jk}}{\partial H_i}\right)_{s,{\bf w}}\ .
\label{pme2}
\end{equation} 
These coefficients describe how strain can induce magnetization and 
how magnetic intensity can induce stress. 
Some implications of this conversion from mechanical energy into electromagnetic 
energy are quite striking. If a solid has an impact on the surface of a 
magnetostrictive material, then the resulting induced electric fields can measure 
the nature of the induced impact stress. 

The piezomagnetic coefficients describe the conversion of phonons 
(mechanical vibrations) into magnons (magnetic energy). Since the magnetization 
yields the current density 
\begin{equation}
{\bf J}_{\rm mag}=c\ curl{\bf M}, 
\label{pme3}
\end{equation} 
Maxwell's equations describe the manner that magnons radiate photons,    
\begin{eqnarray}
{\bf E}({\bf r},t)=-\left(\frac{1}{c}\right)
curl \int \frac{\dot {\bf M}({\bf r}^\prime ,t-R/c)}{R}\ d^3 {\bf r}^\prime , 
\nonumber \\ 
{\rm wherein}\ \ R=|{\bf r} - {\bf r}^\prime|\ \ {\rm and} 
\ \ \frac{\partial {\bf M}({\bf r},t)}{\partial t}
\equiv \dot {\bf M}({\bf r},t).    
\label{pme4}
\end{eqnarray} 
Under conditions of rock crushing\cite{Koshevaya:2008}, the magnetization 
changes give rise in the microcracks\cite{Gruerro:2012} to microwave 
radiation as given by Eq.(\ref{pme4}). Microwaves can in turn give rise 
to accelerated electrons to an energy \begin{math} \gamma mc^2  \end{math}. 
Let us consider this in more detail.

\subsection{Loadstone Fractures \label{lf}}

The formation of microcracks during fracture can be formulated in terms of 
the elastic properties of the 
solid\cite{Landau:1970,Freund:1998,Griffith:1921}. Let 
\begin{math} \sigma_{\rm bond} \end{math} be the fracture stress calculated 
on the basis of chemical bonds. The chemical bond stress can be expressed in 
terms of the elastic properties of the crystal 
\begin{equation}
\sigma_{\rm bond}=\left[\frac{2\cal E}{\pi (1-\nu^2)}\right], 
\label{pme5}
\end{equation}
wherein \begin{math} {\cal E} \end{math} is Young's modulus and  
\begin{math} \nu \end{math} is Poisson's ratio. 
The experimental tensile stress \begin{math} \sigma_F \end{math} 
of a loadstone obeys 
\begin{equation}
\sigma_F \ll \sigma_{\rm bond}   
\label{pme6}
\end{equation}
due to the intrinsic surface tension of the microcrack walls.

For loadstone we have the following numerical estimates;  
\begin{equation}
\sigma_{\rm bond} \sim 10^{12}\ \frac{\rm erg}{\rm cm^3} 
\ \ \ {\rm and}
\ \ \ \sigma_F \sim 10^{10}\ \frac{\rm erg}{\rm cm^3}.     
\label{pme7}
\end{equation}
During the time of microcrack formation, the stress within the empty open 
sliver is determined in order of magnitude by the Maxwell pressure tensor; i.e.
\begin{equation}
\sigma_F \sim \frac{\overline{|{\bf E}|^2}}{4\pi }
\ \ \ \Rightarrow \ \ \ \overline{|{\bf E}|^2}\sim 10^{11}\ {\rm Gauss^2}.      
\label{pme8}
\end{equation}
One may now compute the energy of electrons formed near the microcrack walls 
due to electric field generation. 

\subsection{Electron Energy \label{ee}}

The rate of change of momentum of an electron in an electric field is given by 
\begin{equation}
\frac{d{\rm p}}{dt} = e{\bf E}.
\label{ee1}
\end{equation}
Putting 
\begin{equation}
mc^2 \gamma =\sqrt{c^4m^2+c^2\overline{|{\bf p}|^2}}\ ,
\label{ee2}
\end{equation}
yields  
\begin{equation}
\gamma ^2 = 1+\frac{\overline{|{\bf p}|^2}}{m^2c^2} = 
 1+\frac{e^2\overline{|{\bf E}|^2}}{m^2c^2\Omega^2}\ ,
\label{ee3}
\end{equation}
wherein Eq.(\ref{ee1}) has been invoked 
and \begin{math} \Omega  \end{math} is an effective frequency of electric field 
fluctuations\cite{Widom:2013}
\begin{equation}
\frac{1}{\Omega^2}=\frac{\int |{\bf E}_\omega |^2(d\omega /\omega^2)}
{\int |{\bf E}_\omega |^2 d\omega }\ .
\label{ee4}
\end{equation}
Since 
\begin{equation}
\frac{e}{mc}=1.75882915\times 10^7 \  \left(\frac{1}{\rm Gauss\ sec}\right),
\label{ee5}
\end{equation}
Eqs.(\ref{pme8}), (\ref{ee3}) and (\ref{ee5}) yield 
\begin{equation}
\gamma \sim \frac{3\times 10^{12}/{\rm sec}}{\Omega }\ . 
\label{ee6}
\end{equation}
For the microwave electromagnetic radiation within the microcrack, 
\begin{math} \Omega \sim 10^{10} /{\rm sec} \end{math} so that 
\begin{equation}
\gamma \sim 300 
\ \ \ \Rightarrow 
\ \ \ \gamma mc^2 \sim 150\ {\rm MeV}. 
\label{ee7}
\end{equation}
Such energetic electrons are perfectly capable of inducing the photo-disintegration 
of nuclei. 

\subsection{Electron Deposition \label{ed}}

One can determine the order of magnitude of the number of energetic electrons 
which arrive on the surface area of a microcrack wall during the formation. If 
\begin{math} n_s  \end{math} represents the number per unit area of excess 
electronic charges, then applying Gauss' law to the wall surface yields 
\begin{equation}
4\pi e n_s=E_\perp ,
\label{ed1}
\end{equation}
wherein \begin{math}  E_\perp \end{math} is the component of the electric field 
normal to the surface. Eqs.(\ref{pme8}) and (\ref{ed1}) yield the electronic 
deposition density per unit area 
\begin{equation}
n_s\sim 3\times 10^{13}/{\rm cm^2}
\label{ed2}
\end{equation}
If the yield of nuclear fission events per deposited electron is a few percent, then the 
density of fission events per unit area on a newly formed microcrack wall is given by 
\begin{equation}
\varpi_s\sim  \frac{10^{12}\ {\rm fission \ events}}{\rm cm^2}
\label{ed3}
\end{equation}
This is a sufficient number of fission events for forming visible patches of monolayer 
films that would be observable to the eye.

\section{Conclusions \label{conc}}

The recent experimental work on nuclear radiations from fracturing magnetite, 
\begin{math} [Fe_2^{+++}][Fe^{++}][O_4^{--}]\end{math}, can be largely understood. 
The theory contains an important electro-strong interaction component. 
Giant dipole resonances in the iron and oxygen nuclei 
\begin{equation}
\gamma +\ ^{56}Fe \to \ ^{56}Fe^* \ \ {\rm and}
\ \ \gamma +\ ^{16} O \to \ \ ^{16} O^*    
\label{conc1}
\end{equation}
give rise to a conferable diversity of fission nuclear radiation from the excited 
compound nuclear states \begin{math} \ ^{56}Fe^* \end{math} and  
\begin{math} \ ^{16}O^* \end{math} when rocks containing such nuclei are crushed 
and fractured. Nuclear transmutations are present in the fracture of many 
brittle rocks but they are particularly plentiful for the case of magnetite 
rocks, i.e. loadstones. The fission events nucleus are a consequence of 
photo-disintegration. The electro-strong coupling between electromagnetic fields 
in piezomagnetic rocks are closely analogous to the previously discussed case of 
the fracture of piezoelectric rocks. In both cases, electro-weak interactions can 
produce long wavelength neutrons and neutrinos from energetic protons and electrons 
thus inducing nuclear transmutations. The electro-strong condensed matter coupling 
discussed herein represents new many body collective nuclear photo-disintegration 
effects.

\end{document}